\documentclass{aa}
\usepackage{ffbib}
\usepackage{psfig}
\usepackage{epsfig}
\def\gsim{\;\lower4pt\hbox{${\buildrel\displaystyle >\over\sim}$}\,}
\def\lsim{\;\lower4pt\hbox{${\buildrel\displaystyle <\over\sim}$}\,}

\def \xmm {{\em XMM-Newton}}
\def \src {3C58}

\def \ergsec{\hbox{erg s$^{-1}$}}
\def \hcm {\hbox {\ifmmode $ atom cm$^{-2}\else atom cm$^{-2}$\fi}}
\def \arcmin {\hbox{$^\prime$}}
\def \arcsec {\hbox{$^{\prime\prime}$}}

\def\approxgt{\mathrel{\hbox{\rlap{\lower.55ex \hbox {$\sim$}}
        \kern-.3em \raise.4ex \hbox{$>$}}}}
\def\approxlt{\mathrel{\hbox{\rlap{\lower.55ex \hbox {$\sim$}}
        \kern-.3em \raise.4ex \hbox{$<$}}}}

\begin{document}


\title{The X-ray nebula of the filled center supernova remnant 3C58 and
its interaction with the environment\thanks{Based on observations
obtained with \xmm, an ESA science mission with instruments and
contributions directly funded by ESA Member States and the USA (NASA)}}


      \author{F. Bocchino\inst{1,2}
         \and R. S. Warwick\inst{3}
         \and P. Marty\inst{4}
         \and D. Lumb\inst{1}
         \and W. Becker\inst{5}
         \and C. Pigot\inst{6}
}
\offprints{F. Bocchino (fbocchin@estec.esa.nl)}

\institute{
       Astrophysics Division, Space Science Dept. of ESA, ESTEC,
              Postbus 299, 2200AG Noordwijk, The Netherlands
\and
       Osservatorio Astronomico di Palermo, Piazza del Parlamento 1,
       90134 Palermo, Italy
\and
       Department of Physics and Astronomy, University of Leicester,
       Leicester LE1 7RH, U.K.
\and
       Institut d'Astrophysique Spatiale, Campus Universite Paris-Sud, 
       F-91405 Orsay Cedex, France
\and
       Max-Planck-Institut f\"ur extraterrestrische Physik,
       Giessenbachstarsse 1, D-85740 Garching, Germany
\and
       DAPNIA / Service d'Astrophysique, CEA/Saclay, F-91191 Gif-Sur-Yvette, 
       CEDEX, France
}

\date{Received 5 December 2000; Accepted: 30 January 2001}

\abstract{An \xmm\  observation of the plerionic supernova remnant 3C58
has allowed us to study the X-ray nebula with unprecedented detail.  A
spatially resolved spectral analysis with a resolution of 8\arcsec\
has yielded a precise determination of the relation between the
spectral index and the distance from the center. We do not see any
evidence for bright thermal emission from the central core. In contrast
with previous ASCA and {\em Einstein} results, we derive an upper limit
to the black-body 0.5--10 keV luminosity and emitting area of
$1.8\times 10^{32}$ \ergsec and $1.3\times 10^{10}$ cm$^2$,
respectively, ruling out emission from the hot surface of the putative
neutron star and also excluding the ``outer-gap" model for hot polar
caps. We have performed for the first time a spectral analysis of the
outer regions of the X-ray nebula, where most of the emission is still
non-thermal, but where the addition of a soft (kT=0.2--0.3 keV) optically
thin plasma component is required to fit the spectrum at $E<1$ keV.
This component provides 6\% of the whole remnant observed flux in the
0.5--10.0 keV band. We show that a Sedov interpretation is incompatible
with the SN1181-3C58 association, unless there is a strong deviation
from electron-ion energy equipartition, and that an origin of this thermal
emission in terms of the expansion of the nebula into the ejecta core
nicely fits all the radio and X-ray observations.   }

\maketitle

\markboth{F. Bocchino et al.}
{X-rays from \src}

\section{Introduction}

3C58 is a beautiful example of a filled-center (or plerionic) supernova
remnant (SNR), probably associated with the historical supernova event
in 1181 A.D. (\cite{ste71}). This object has always received much
attention because, on the one hand, it shows some characteristics similar to
the Crab SNR, while, on the other hand, it seems very different to the
Crab itself. For instance, 3C58 has a compact ($10^\prime\times 6^\prime$)
elliptical morphology with a very bright core (\cite{ra88}) and linear
size similar to the Crab; \cite*{fm93} have reported a wisp-like
elongated structure at $2.6\arcsec$ from the core, which has been
observed also in the Crab and which is probably associated with the
pulsar wind termination shock.  However, unlike the Crab, there is no
clear evidence of a pulsating point source located in the center, as
would be expected, since the morphology strongly suggest that the
nebula is powered by a spinning neutron star. Despite considerable effort,
pulsations have not been detected so far in either radio (see e.g.
\cite{llc98}), nor in X-rays (\cite{hbw95}, H95 hereafter). Moreover,
the X-ray to radio flux ratio ($f_X/f_r$) of 3C58 is 100 times lower than that
of the Crab (H95), its spectral break occurs at 50 GHz (300 times less
than the break of the Crab, \cite{gs92}) and its radio luminosity is
increasing instead of decreasing as expected (\cite{gre87}). These
remarkable differences are also seen in other plerions, and
\cite*{wsp97} have proposed the introduction of a new sub-class of
plerions, the ``non Crab-like plerions", of which 3C58 can be
considered the prototype. For these objects, a non-standard evolution of
the pulsar can be invoked, but the details are not yet clear.

It is therefore very important to investigate the physical properties
which render 3C58 so peculiar, in order to put this object and its
sub-class in the right perspective. In particular, the detection or
non-detection of the central source is obviously a key point.
\cite*{bhs82} reported the presence of a compact X-ray source in 3C58
from their Einstein HRI observation, about 10\arcsec\  in extent, and
contributing to 15\% of the detected X-ray flux. Later, \cite*{hbw95}
revisited the X-ray emission of 3C58 using ROSAT HRI data, confirming
the compact source and favored a model in terms of hot polar caps
to explain the emission. However, it has not
been possible so far to take an X-ray spectrum of the source to study
it, and to understand if it is really a point source or an enhancement
of the pulsar nebula. \cite*{tsk00} have pointed out that the inclusion
of a black-body component in the fit of the ASCA GIS+SIS data of 3C58
yields an improvement of the $\chi^2$. They claim that the best-fit
black-body component is responsible for $\sim 7$\% of the unabsorbed
flux in the 0.5-10 keV of the whole remnant, and that it is the
spectral signature of the central source.

It is also very important to assess the presence of a shell around the
pulsar nebula, for it may give compelling constraints on the age of the
remnant, the shock velocity and the density of the environment.  In the
case of the $\sim 800$ yr old 3C58 (as other plerions as well), it is
expected that the main shock will encounter the stellar ejecta and/or
the interstellar medium (ISM) giving rise to a limb brightened shell.
However, no sign of a shell has been detected at centimeter wavelengths
at distances greater than $5\arcmin$ from the core
(\cite{ra85}). However, \cite*{ra88} have imaged the faint outer
emission of the nebula at a distance between 2\arcmin\  and $4\arcmin$
and have noticed limb brightening at several
locations. In this
paper, we present a study of the \xmm\  data of 3C58 obtained during the
Calibration and Performance Verification (Cal/PV) phase of the mission.
We show that a stringent upper-limit can be imposed on the
presence of thermal emission from a central point source, and we also find
evidence for a thermal outer shell. The implications of these findings
for current models of 3C58 are also discussed.

\section{Data analysis}

\subsection{Observations}

3C58 has been observed as part of the Cal/PV phase of the \xmm\ 
satellite (\cite{jla00}). In this paper, we focus on the 25 ks
on-axis observation performed during orbit 47, on March 12th 2000.
Data from the two MOS (\cite{taa00}) cameras and the PN (\cite{sbd00})
camera have been used. MOS and PN cameras are CCD arrays which collect
X-ray photons between 0.1 and 15 keV and have a field of view of
$30^\prime$. The pixel size is $1.1\arcsec$ and $4.1\arcsec$ for MOS
and PN respectively, and this should be compared with the mirror PSF
width of $6\arcsec - 15\arcsec$ FWHM--HEW. The data have been acquired
with the medium filter and in full image mode, and therefore the
temporal resolution is low, 2.5 s and 73 ms for MOS and PN,
respectively. The worse spatial resolution of the PN is compensated by
its greater sensitivity, on the average 20-30\% more than the combined
two MOS.

The Standard Analysis System (SAS) software we have used (version 5.0
alpha, xmmsas-20001011-1559) takes care of most of the required events
screening. However, we have further screened the data to eliminate some
residual hot pixels and occasional background enhancement due to
intense flux of soft protons in the magnetosphere. In particular, we
have extracted the background light-curve from a region free of
sources, and we have identified time intervals of unusually high count
rates (more than twice the ``quiescent" background rate) and removed
them from subsequent analysis. Moreover we have also selected events
with value of the PATTERN pipeline assigned keywords between 0 and 12.
The exposure time of the screened observations is 16 ks for the MOS and
12 ks for the PN detectors.



\subsection{X-ray morphology}

\begin{figure}
  \centerline{\psfig{file=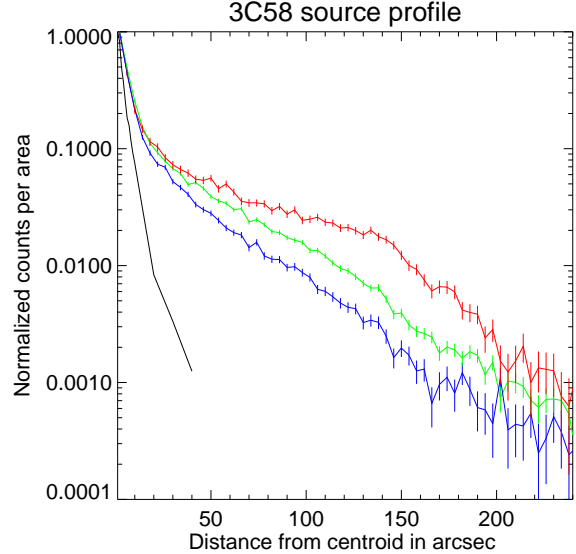,width=8cm}}
  \caption{MOS1+2 source profile, centered at $2^h05^m38.0^s$
  $+64^d49^m37^s$ (J2000). Soft, medium and hard energy band as
  defined in Fig. \protect\ref{mos_imh} are represented in red, green
  and blue, respectively. The solid black line is the Point Spread
  Function (PSF) of the mirror module 3 at 1.5 keV (\protect\cite{abh00})}
  \label{mos_prof}
\end{figure}

\begin{figure}[!ht]
  \centerline{\psfig{file=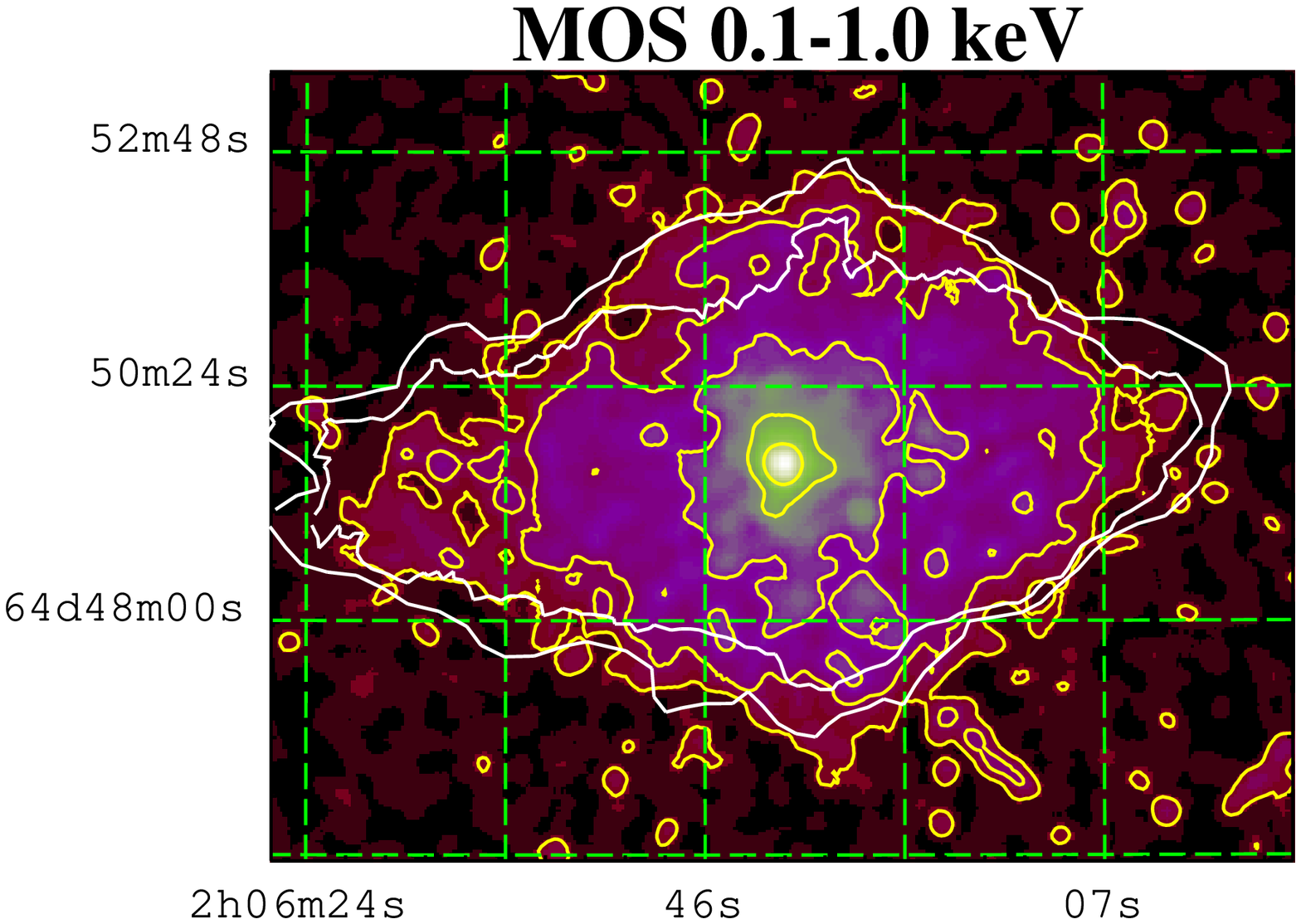,width=9.0cm}}
  \centerline{\psfig{file=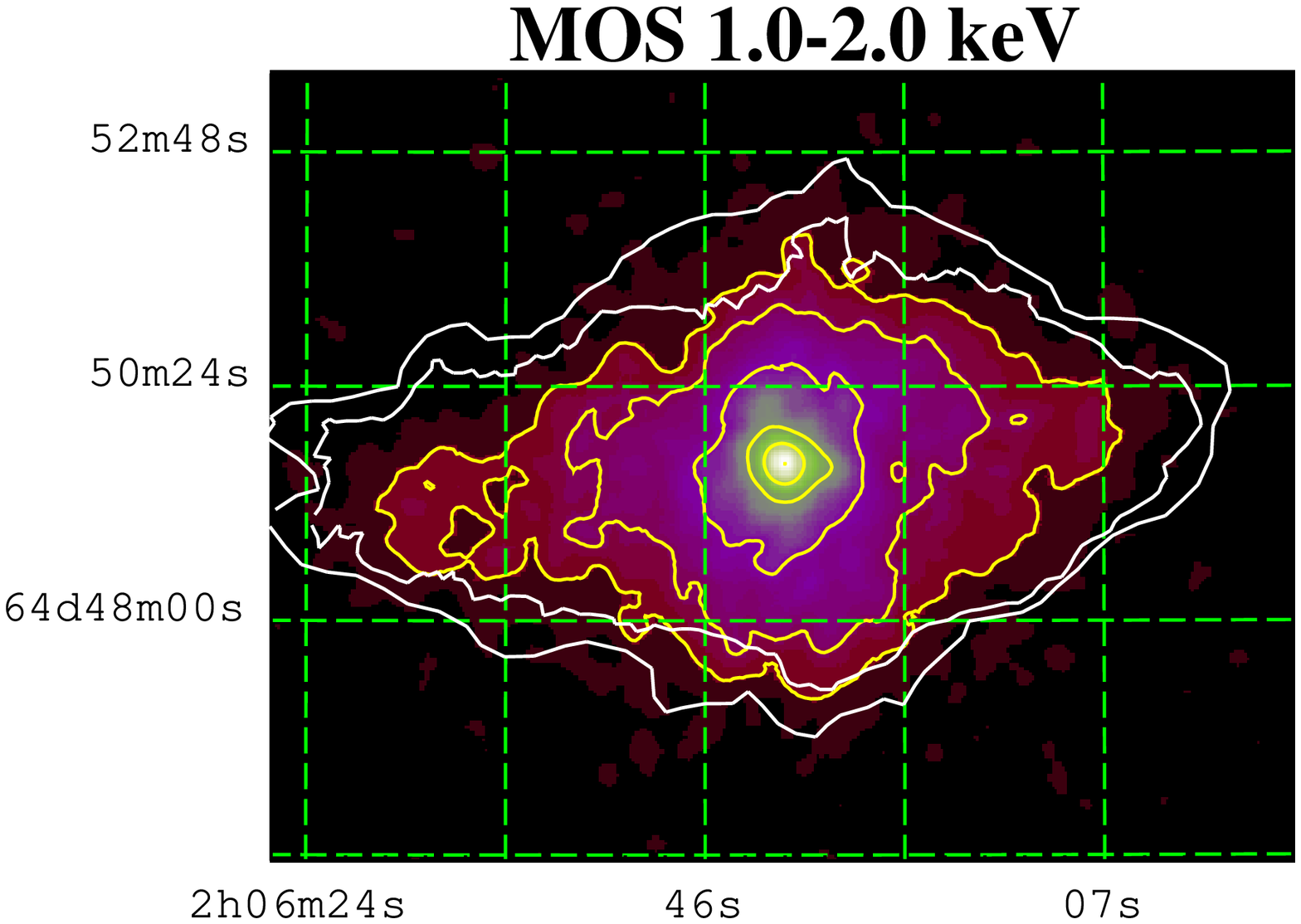,width=9.0cm}}
  \centerline{\psfig{file=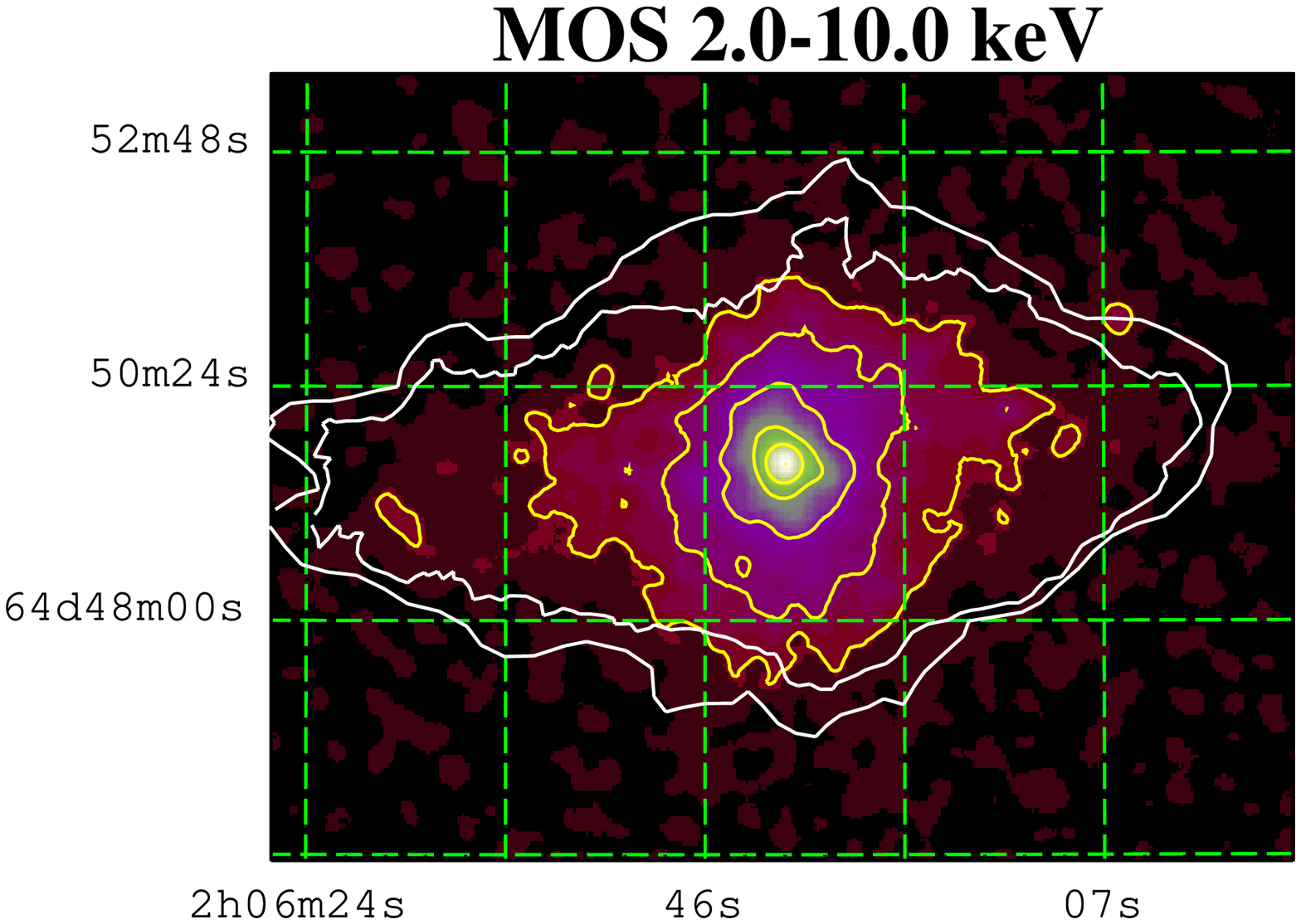,width=9.0cm}}
  \caption{MOS1+2 images binned at pixels of $2\arcsec$ and smoothed
  using a gaussian with $\sigma=3$ pixel to enhance the morphology of
  the faint nebula edge. {\em Upper panel:} soft image (0.1--1.0 keV).
  {\em Center panel:} medium energy image (1.0--2.0 keV). {\em Lower
  panel:} hard image (2.0--10.0 keV).  Overlaid are contours
  representing the weakest ($3\sigma=0.15$ Jy beam$^{-1}$) 1446 MHz
  radio contour and the one corresponding to a flux density 4 times
  higher (both in white, from \protect\cite{ra88}), and six X-ray
  contours (in yellow, from 1/100 of the peak to the peak value,
  logarithmically spaced).}

  \label{mos_imh}
\end{figure}

In Fig.  \ref{mos_prof}, we report the 3C58 profile as seen by the
MOS1+2 computed in 65 annuli with $\Delta r = 4\arcsec$ up to a
distance of $\sim 4\arcmin$ from the centroid, located at
$2^h05^m38.0^s$ $+64^d49^m37^s$ (J2000, $\pm 1\arcsec$).  The
background was collected from a ring with $\Delta r = 30\arcsec$
located at $4.5\arcmin$ from the center, and has been chosen to fall
entirely in the central MOS chip. The comparison with the expected
mirror PSF, also shown in Fig. \ref{mos_prof} indicates that the
central source is extended. The bright core region can be defined up to
$\sim 50\arcsec$ from the center; Further out, the slope changes and the
behavior of the soft curve and hard curve is different. The hard curve
declines with the same slope down to the limit of the radio nebula at
$\sim 200\arcsec$, while the soft curve is clearly flatter than the
hard one, with a sudden change of slope occurring at $140\arcsec$.

\begin{figure}
  \centerline{\psfig{file=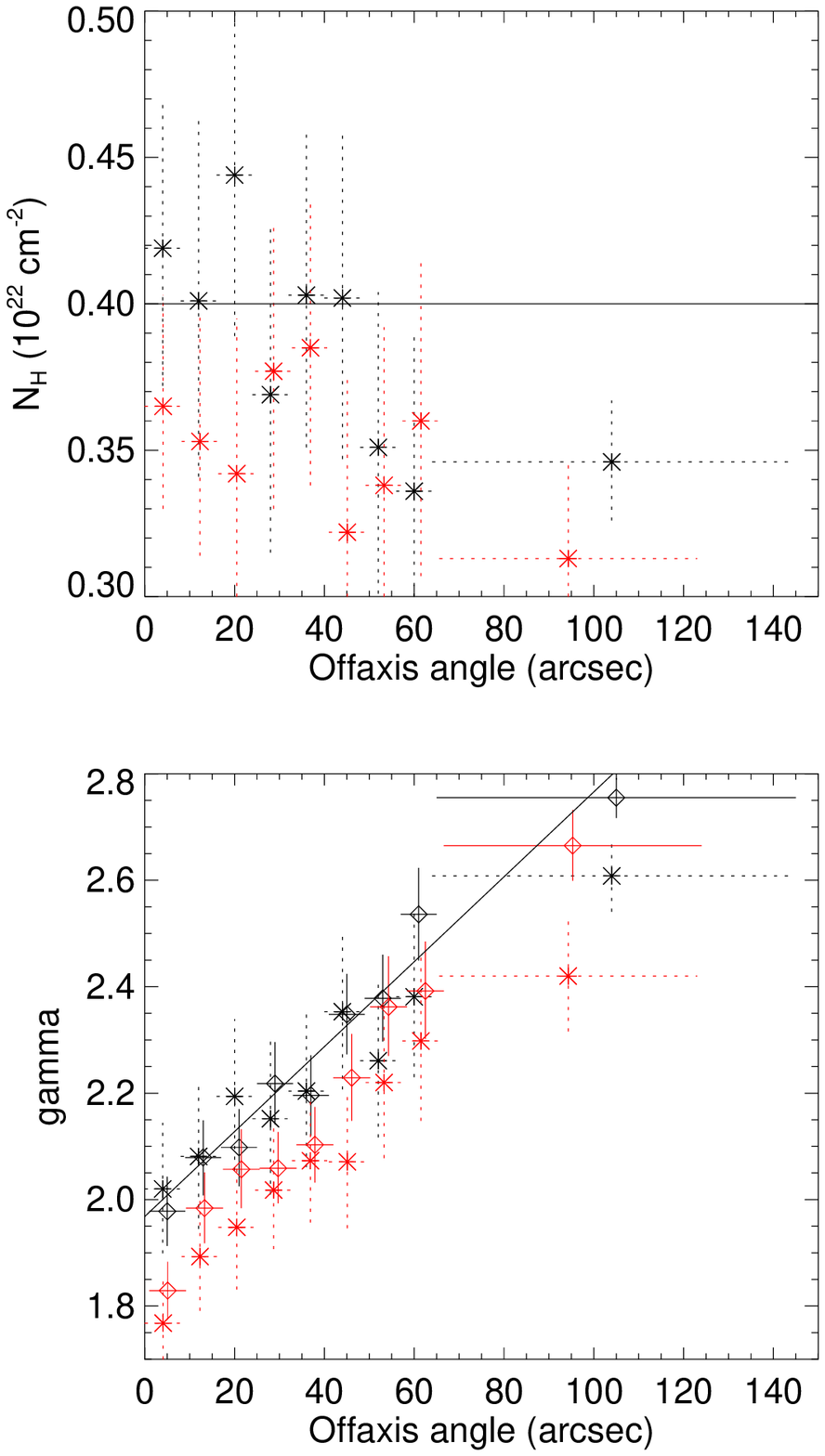,width=9.5cm}}
  \caption{Spatially resolved spectral analysis with a power-law model
  of the X-ray nebula associated with 3C58. We show the results
  obtained on the 8 annuli and the ``edge" regions (the last point) as
  defined in the text. Black is MOS, red is PN, dotted bars are fit
  with free $N_H$, solid bars are fit with $N_H$ fixed at $4\times
  10^{21}$ cm$^{-2}$. Regions are centered at $2^h05^m38.0^s$
  $+64^d49^m37^s$ (J2000). A linear fit to the MOS points corresponding
  to fits with $N_H$ fixed at $4\times 10^{21}$ cm$^{-2}$ is shown in
  the lower panel.}
  \label{rgamma} 
\end{figure}


The core is elongated in the N-S direction, with a FWHM of
$13.5\pm1.5\arcsec$ in this direction ($9.0\pm1.5\arcsec$ in the E-W
direction), in agreement with the ROSAT HRI results presented by H95.
The centroid of the MOS1 emission is located 4\arcsec\ and 8\arcsec\
south of the ROSAT and Einstein positions.  The data confirm that the
radio wisp observed by \cite*{fm93} is located to the East of the X-ray
peak, at 5\arcsec\ from the the MOS1 centroid.  However, the
uncertainties in real attitude reconstruction, which is still not fully
implemented, may give an uncertainty in absolute position determination
as large as $\sim 5\arcsec$.

Fig. \ref{mos_imh} shows the summed MOS1+2 images of 3C58 in three
different bands, namely 0.1-1 keV, 1--2 keV and 2--10 keV bands, along
with the weakest 1446 MHz radio contour at 0.15 Jy beam$^{-1}$ and the
contour at 0.6 Jy beam$^{-1}$ of Fig. 3 in \cite*{ra88} (both in
white in Fig. \ref{mos_imh}). A small separation of the two contours
(e.g. in the SW) indicates limb brightening and therefore confinement
in the source. In addition, we also shown the X-ray contours (in
yellow), and it is clear that there is correspondence between the radio
contours and the X-ray emission of the soft image. In particular, the
weakest radio contour matches the outer edge of the weak extended X-ray
emission, especially at the north and south edges. Fig.  \ref{mos_imh}
also shows that the size of the nebula decreases as the energy
increases, an effect also reported by \cite*{tsk00}.

\subsection{Spectral analysis}

The high spatial and spectral resolution of the instrumentation aboard
the \xmm\  satellite allows us for the first time to perform spatially
resolved spectral analysis of the X-ray nebula associated to 3C58. To
this end, we extracted spectra from 8 concentric annuli with $\Delta r
= 8\arcsec$ centered at the same position of the X-ray profiles of Fig.
\ref{mos_prof}, covering the core emission up to 1.1\arcmin\  from the
center. Given the uncertainties in present calibration of the MOS and
PN cameras, and the energy dependence of the vignetting correction
above 5 keV, we have restricted our spectral analysis to photons in the
0.5--5 keV energy band.  The spectra have been background subtracted
using the same background region introduced in the previous subsection,
and have been rebinned to ensure that a minimum of 30 counts are
present in each energy channel. As for response matrices and effective
area files, we have used the latest version of standard MOS and PN
matrices provided by the calibration team
(mos1\_medium\_all\_qe17\_rmf3\_tel5\_15 and epn\_fs20\_sY9\_medium).
We have summed the spectra extracted with MOS1 and MOS2, and we have
also rescaled the response matrix to reflect this operation\footnote{We
will refer hereafter to the summed spectra as simply the MOS
spectra.}.  Besides the 8 annuli, we have also defined a region (the
"edge" region hereafter), represented by the union of two ellipses
matching the outer X-ray edge of the nebula\footnote{The E-W and N-S
ellipses have large and small semi-axis of $3.3\arcmin\times 1.1\arcmin$
and $1.7\arcmin\times 1.1\arcmin$, respectively.}, minus a circle with
same radius as the 8th annulus. This region is particularly suited for
the study of the X-ray emission coming from the outermost fringes of
the X-ray and radio nebula.

We have used three different emission models to fit the 3C58 data,
namely a power-law model, a power-law model plus the optically thin
plasma model of \cite*{mgo85} with Fe L calculation of \cite*{log95},
pl+{\sc mekal} hereafter, and a power-law model plus a
black-body spectrum, pl+bbody hereafter. These three models encompass
what we could possibly expect from an X-ray nebula, the last two
representing eventual contributions from a thermal shell (as in the
case of known plerion-composite SNR), and from a compact source in the
center as pointed out by \cite*{hbw95}. The temperature of the pl+{\sc
mekal} model and of the pl+bbody models have been constrained in the
0.1--10 keV and in the 0.1--2.0 keV, respectively. Abundances are those
of \cite*{ag89}. All the models have been modified by the interstellar
absorption according to cross-sections of \cite*{mm83}, where we have
let the equivalent hydrogen column density $N_H$ vary. Since we have
noted that the non-thermal component provides most of the flux in the
\xmm\  bandwidth, and that the residual thermal components of the model
pl+{\sc mekal} and pl+bbody are correlated with the value of $N_H$, we
have also performed a set of fittings fixing the $N_H$ value to
$4\times 10^{21}$ cm$^{-2}$, which is compatible with previous
estimates (\cite{hbw95},\cite{tsk00}) and it is also consistent with
the result we obtained letting it vary.

\begin{table*}
\caption{Goodness of MOS fits to the 8 annuli of 3C58. The absorption
is fixed to $4\times 10^{21}$ cm$^{-2}$.  For each model, we report the
value of $\chi^2$, degrees of freedom (dof), null hypothesis
probability. For the pl+{\sc mekal} and pl+bbody model, we also report
the value of the best-fit temperature.  The last column is the
unabsorbed flux in the 0.5--10.0 keV in units of $10^{-13}$ erg
cm$^{-2}$ s$^{-1}$; its statistical error is $\pm 0.1$.}
\medskip
\begin{tabular}{lccccccccc} 
\hline\noalign{\smallskip}
Reg. & \multicolumn{2}{c}{Power-law} & \multicolumn{3}{c}{Pow+{\sc mekal}} & \multicolumn{3}{c}{Pow+bbody} \\
  & $\chi^2/dof$   & Prob. & kT & $\chi^2/dof$   & Prob. & kT & $\chi^2/dof$   & Prob. & flux \\
\noalign{\smallskip\hrule\smallskip}
1 &  74/73 &   46 & $1.0(>0.4)$ & 71/71 &   48 & $0.2(>0.1)$ & 73/71 &   40 & 15.3 \\
2 &  69/65 &   35 & $6.0(>0.1)$ & 67/63 &   35 & $0.9(>0.1)$ & 68/63 &   30 & 13.0 \\
3 &  65/63 &   41 & $0.6(>0.1)$ & 65/61 &   34 & $0.5\pm0.3$ & 62/61 & 44 & 11.4 \\
4 &  52/59 &   71 & $0.1(>0.1)$ & 51/57 &   70 & $0.1(>0.1)$ & 51/57 &   69 & 12.3 \\
5 &  54/61 &   72 & $0.7(>0.1)$ & 52/59 &   72 & $0.2(>0.1)$ & 53/59 &   69 & 12.0 \\
6 &  52/59 &   74 & $0.1(>0.1)$ & 50/57 &   72 & $0.6(>0.1)$ & 50/57 &   72 & 10.9 \\
7 &  96/60 & 0.2  & $0.2(<0.25)$& 91/58 &  0.4 & $0.8^{+0.2}_{-0.1}$ & 91/58 & 0.4 & 10.7 \\
8 &  79/58 &  3   & $0.3^{+0.2}_{-0.1}$ & 71/56 & 8 & $0.8^{+0.3}_{-0.1}$ & 73/56 & 7 & 10.6 \\

\noalign{\smallskip}
\hline
\end{tabular}
\label{gof}
\end{table*}

\subsubsection{The $\gamma$ vs. radius relation}

Fig. \ref{rgamma} shows the best-fit value of the absorption and of the
power-law photon index ($\gamma$) as a function of the distance from
remnant center, obtained with a fit to a power-law emission model only. 
The data clearly show the effect of synchrotron
burn-off of high energy electrons as the radius increases. This effect
has also been observed in G21.5-0.9 both with Chandra (\cite{scs00})
and \xmm\  (\cite{wbb00}) and it is related to inhomogeneity in
the particle distribution inside the plerion
nebula. The straight line in the lower panel of Fig. \ref{rgamma}
represents the linear best-fit to the $\gamma - r$ relation,
$\gamma=A+Br$ with $A=1.97\pm 0.03$ and $B=8.0\pm 0.6\times 10^{-3}$
and $r$ is in arcseconds. The MOS fits to the single power-law
reported in Fig. \ref{rgamma} are statistically acceptable\footnote{We
regard a fit as statistically acceptable when the null hypothesis
probability computed from the reduced $\chi^2$ and number of degrees of
freedom is greater then 5\%.} from ring 1 to ring 6, while not
acceptable in rings 7-8 and in ``the edge" region. This is true for
both $N_H$ free and $N_H$ fixed fits, and Table \ref{gof} reports the
$\chi^2/dof$ values of the fits. It is interesting to note that the
$N_H$ value, when left free to vary, is significantly lower than the average
value of $4\times 10^{21}$ cm$^{-2}$ for the outer nebula regions (Fig.
\ref{rgamma}).  Moreover, the best-fit $\gamma$ of the shell is off the
trend dictated by fit to the spectra of the rings when $N_H$ is left
free to vary, while it shows lower deviation in the fit with $N_H$
fixed. If the X-ray emission of the outer rings is dominated by the
non-thermal component of the plerion, we do not expect significative
variation of the absorption, and the data seem to confirm that a fixed
$N_H$ may be more appropriate.  However, the fact that fits of the
outer rings and ``edge" data are less acceptable than fits to inner
rings data strongly suggests a contribution from other components.

\subsubsection{Additional thermal components}

\begin{figure*}
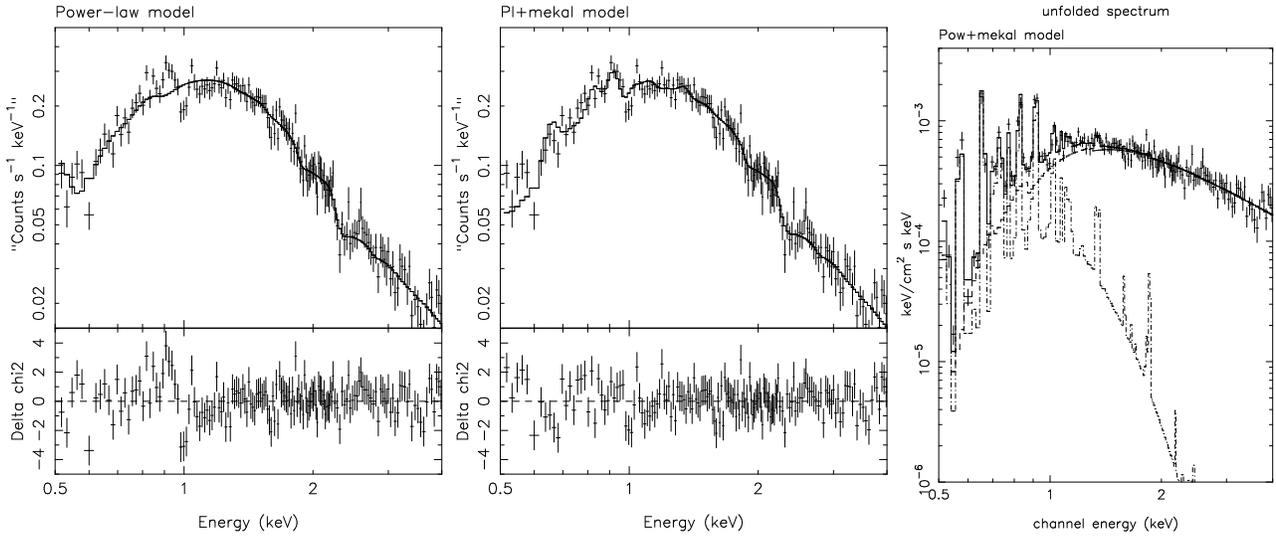

  \centerline{\hbox{
     \psfig{file=FIGURES/shellpow.ps,width=7cm,angle=-90}
     \psfig{file=FIGURES/shellpowmekal.ps,width=7cm,angle=-90}
     \psfig{file=FIGURES/shellunf.ps,width=7cm,angle=-90}
  }}
  \caption{A close up of MOS spectrum for the ``edge" region. {\em
  Left panel:} the fit with a simple power-law model does not well
  represent the spectrum below 1 keV. {\em Center panel:} the pl+{\sc
  mekal} model with kT=0.2 keV improves the description of the data.
  The spectral region near the oxygen edge (between 0.4 and 0.6) is know to
  have some residual calibration problem.
  {\em Right panel:} Unfolded best-fit pl+{\sc mekal} model. Dot-dashed
  line represents the thermal component, which shows emission lines and
  falls off rapidly above 1 keV, where the power-law component
  dominates.}
  \label{shellpha}
\end{figure*}

\begin{figure}
  \centerline{\psfig{file=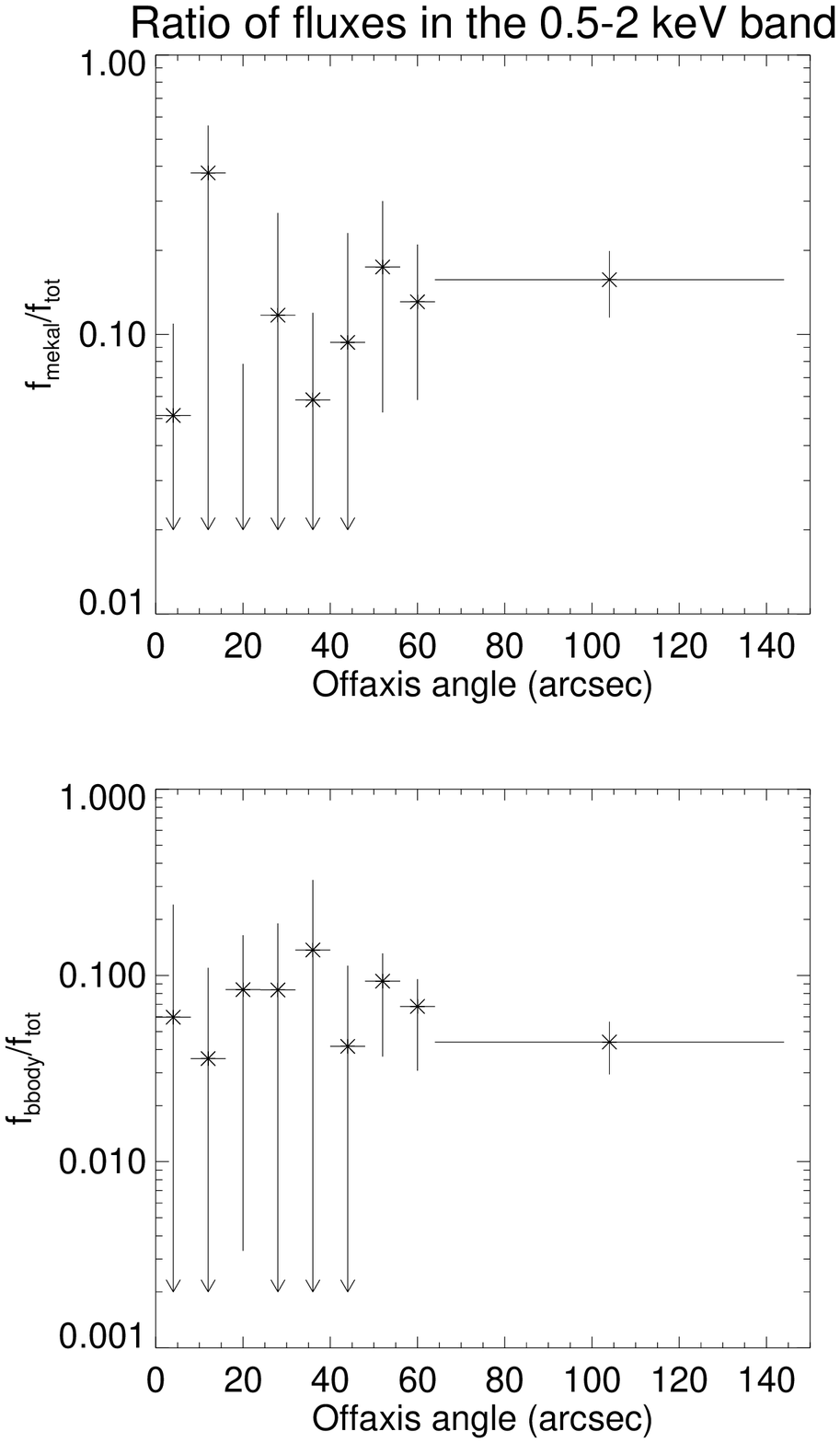,width=9.5cm}}
  \caption{Same as in Fig. \protect\ref{rgamma}, but here we report the
  ratio between the 0.5--2.0 keV flux due to the additional component
  ({\sc mekal} in the upper panel, and bbody in the lower panel) to the
  total 0.5--2.0 keV flux versus the distance from the center. Arrows
  indicate upper limits. Goodness of fits are reported in Table
  \protect\ref{gof}.}
  \label{ratio}
\end{figure}

In order to assess the presence of any additional emission from a central
compact source (as suggested by H95) and from any thermal shell, 
we now consider the results of fits with pl+{\sc mekal} and pl+bbody models.
The inclusion of
the additional component ({\sc mekal} or bbody) does not increase the
null hypothesis probability above 5\%, except for ring 8. This is also
reported in Table \ref{gof}, in which the reader finds the data needed
to evaluate the goodness of the fits we have performed.
However, it should be noted that the inclusion of the {\sc mekal}
component leads to a reduction of $\chi^2$ (and therefore to a better fit
in a relative way) for the outermost rings, and a dramatic reduction in
case of the ``edge" region, specially for MOS. This is also shown in
Fig. \ref{shellpha}, where we show the MOS ``edge" spectrum along with
its power-law only and pl+{\sc mekal} best-fit model and residuals, and
in Table \ref{shellres} where we summarize the results of spectral
fitting to the ``edge" spectrum.
In Fig.  \ref{ratio}, we report the ratio of the flux of the second
component ($f_{\rm mekal}$ from the pl+{\sc mekal} fit in the upper
panel, and $f_{\rm bbody}$ from the pl+bbody fit in the lower panel) to
the total flux of ring 1-8 and the ``edge" region.

The fit with an additional black-body component to ring 1 places an
upper limit to the unabsorbed flux due to this component of 20\% of the
total 0.5--2.0 keV flux in this region (i.e. $1.5\times 10^{-13}$ erg
cm$^{-2}$ s$^{-1}$, or a luminosity $L_X=1.8\times 10^{32}D^2_{3.2}$
erg s$^{-1}$, where $D_{3.2}$ is the distance in units of 3.2 kpc,
which is the most reliable value according to \cite{rgk93}) and 10\% of
the total 0.5--10.0 keV flux of this region (i.e. $1.6\times 10^{-13}$
erg cm$^{-2}$ s$^{-1}$, same $L_X$ as before). Since the whole remnant
0.5--10 keV flux is $1.6\times 10^{-11}$ erg cm$^{-2}$ s$^{-1}$, the
upper limit corresponds to 1\% of the remnant flux, significantly below
the 7\% found by \cite*{tsk00}.  Fig. \ref{ratio} also shows that
spectral fittings of rings 2-6 yields only upper-limits to the presence
of an additional {\sc mekal} or black-body thermal emission.

On the other hand, the situation is different for ring 8 and the
``edge" region (Table \ref{gof} and \ref{shellres}), where the addition
of the thermal component yields a significant decrement of the
$\chi^2$ (according to an F-test) and the thermal flux contribution to
the 0.5--2.0 total flux is between 5\% and 30\%. In particular, the
pl+{\sc mekal} fit in the ``edge" region suggests that between 10 and
20\% of the 0.5--2.0 keV flux from this region ($5-9\times 10^{-13}$
erg cm$^{-2}$ s$^{-1}$, or $L_X=5.8-10.0\times 10^{32}D^2_{3.2}$ erg
s$^{-1}$) is due to the thermal component (8--14\% if we consider the
0.5--10 keV band).  If compared to the whole remnant 0.5--10 keV flux,
the thermal soft excess yields a contribution of 3\%--6\%.
It is important to note that in this region the pl+{\sc mekal} model
is to be preferred over the pl+bbody model (Table \ref{shellres}).

\begin{table}
\caption{PN and MOS fit results of the ``edge" region of 3C58. All
 the fits yield an absorbed flux of $6.1\times 10^{-12}$
 erg cm$^{-2}$ s$^{-1}$ in the 0.5--10 keV band.}
\medskip
\begin{tabular}{lcccc} 
\hline\noalign{\smallskip}
Model & $N_H$ & $\gamma$ & kT & $\chi^2/dof$ \\
      & cm$^{-2}$ &      &  keV \\
\noalign{\smallskip\hrule\smallskip}
  \multicolumn{5}{c}{MOS1+2} \\
pow-law & $3.5\pm 0.2$ & $2.61\pm 0.06$ &       -      & 245/170 \\
pl+{\sc mekal} & $7.3^{+0.6}_{-0.9}$ & $2.89\pm 0.08$ & $0.18\pm 0.01$ & 202/168 \\
pl+bbody & $5.5^{+0.5}_{-0.8}$ & $2.75\pm 0.06$ & $<0.11$ & 231/168 \\
pow-law & $4.0$ & $2.75\pm 0.04$ &        -       & 261/171 \\
pl+{\sc mekal} & $4.0$ & $2.61\pm 0.05$ & $0.25\pm 0.03$ & 226/169 \\
pl+bbody & $4.0$ & $3.07\pm 0.07$ & $0.78\pm 0.08$ & 243/169 \\
\hline\noalign{\smallskip}
  \multicolumn{5}{c}{PN} \\
pow-law & $3.1\pm 0.3$ & $2.42\pm 0.11$ &       -      & 113/138 \\
pl+{\sc mekal} & $7.6^{+0.4}_{-2.5}$ & $2.71^{+0.15}_{-0.11}$ & $0.18\pm 0.03$ & 107/136 \\
pl+bbody & $5.2^{+1.0}_{-1.7}$ & $2.55^{+0.11}_{-0.16}$ & $<0.13$ & 109/136 \\
pow-law & $4.0$ & $2.66\pm 0.06$ &        -       & 130/139 \\
pl+{\sc mekal} & $4.0$ & $2.47\pm 0.10$ & $0.24\pm 0.06$ & 110/137 \\
pl+bbody & $4.0$ & $2.46\pm0.10$ & $<0.13$ & 110/136 \\


\noalign{\smallskip}
\hline
\end{tabular}
\label{shellres}
\end{table}

\subsubsection{PSF effects}
We have also investigated whether our results are affected by any effect
related to the Point Spread Function (PSF) of the \xmm\  mirror. Since
only 55\% of the photons of a point source are contained within a circle
of 8\arcsec\  radius, we expect a certain amount of scattering outside
the rings. However, this effect is not likely to modify our
conclusions.  In the central ring, for instance, the strongly peaked
morphology keeps the contamination from ring 2 low. Moreover, the
scattering can only smooth the $\gamma-r$ relation in Fig.
\ref{rgamma}, so the observed slope is, strictly speaking, a lower
limit to the real slope. Finally, in order to verify whether the soft
component detected at large radii is an artifact of the PSF, we have
followed this procedure: we have fitted the summed spectrum of rings
1-8 with a power-law model, and we have found $N_H=3.9\times 10^{21}$
cm$^{-2}$ and $\gamma=2.21$; then, we have fitted the spectrum of
the ``edge" region with the best-fit model found for ring 1-8 with
fixed $N_H$ and $\gamma$, plus a second absorbed power-law model with
free $N_H$ and $\gamma$. The normalizations of the two models were left
free to vary. We have found that the best-fit normalization of the
``fixed" power-law is zero, suggesting that negligible contamination
from the spectrum of the ring 1-8 is observed in the ``edge" spectrum.

\subsection{Timing analysis}

In order to search pulsed X-ray emission from the central source which
is powering 3C58, we have calculated the power spectrum density of the
time series of events of the central ring used for spectral analysis.
Unfortunately, the time resolution of the EPIC cameras when operated in
full image mode is low, and we were not able to sample frequencies
above 1 Hz. For timing analysis, we have collected MOS1 and PN photons
in ring 1 between 0.5 and 10 keV, and we have analyzed the two
independently. Furthermore, we have also analyzed a more restricted
energy range, 2--10 keV.

None of the power spectra show features significant at the 99\%
confidence level. The corresponding MOS1 upper limits to relative
amplitude of a sinusoidal pulsed signal in the $5\times 10^{-3}-0.1$ Hz
is 6.2\% and 10.0\% for the 0.5--10 keV and 2--10 keV respectively, while
the PN upper limits in the $10^{-2}-1$ Hz are 2.0\% and 3.1\% in the broad
and hard band, respectively.

\section{Discussion}

The good spatial and spectral resolution \xmm\  data on 3C58  has
allowed us to perform a detailed spatially resolved spectral analysis of
this remnant, and in particular, to address the topic of the spectra of
the compact core at its center and the presence of an X-ray shell.

\subsection{The compact core}

\begin{figure}
  \centerline{\psfig{file=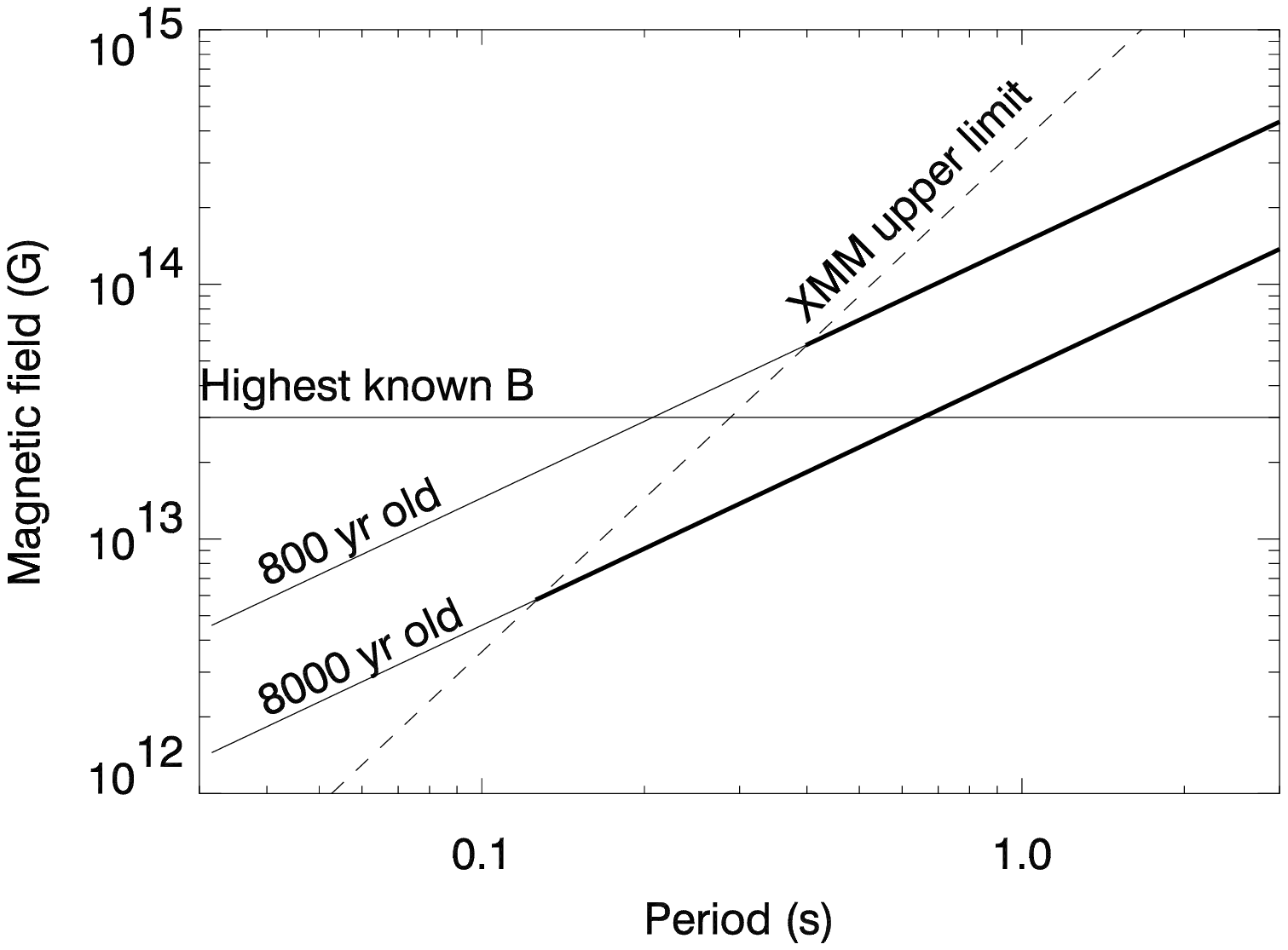,width=10cm}}
  \caption{The plot shows that allowed regions of the $B-P$ plane
  according to the classical model for pulsars (e.g.
  \protect\cite{mt77}) in solid lines (assuming the initial period was
  much less than the present period). We show the case for an 800 yr
  old pulsar (i.e.  assuming that 3C58 is related to SN1181) and for a
  pulsar with an age a factor 10 larger. The dashed line represent the
  $B-P$ values allowed by the ``outer-gap" polar heated caps model of
  \protect\cite*{chr86} and our derived upper limit to the black-body
  luminosity of the central source. The intersection of the
  observational and theoretical curves gives the allowed $B-P$ values
  (thicker part of the isochrone lines). If 3C58 is the remnant of
  SN1181, the model would provide an unreasonably high $B$. Older
  pulsars and/or low rotational energy losses may yield more reasonable
  $B$ values.}
  \label{h95rev}
\end{figure}

The possible emission mechanisms for the compact core of 3C58 have been
reviewed by H95, and a preference has been given to a thermal model for
the emission from hot polar caps, mostly on the basis of a review of
ROSAT and Einstein HRI data. \cite*{tsk00} have confirmed the detection
of an additional component in the remnant spectrum, which they have
modeled with a black-body of kT=$0.40-0.45$ keV and a flux of
$1.5\times 10^{33}D^2_{3.2}$ \ergsec\  in the 0.5--10 keV band. While
the \cite*{tsk00} flux estimate is within 20\% of the H95 one, we have
found in ring 1 an upper limit to the flux of the kT=0.25 keV
black-body component a factor of 10 less than the H95 estimate. The
inferred upper limit on the emitting area is $1.3\times 10^{10}$ cm$^2$
(corresponding to a radius of $\sim 650$ m), more or less the value
found by \cite*{tsk00}, and this is expected since our best-fit
temperature is a factor of two less than the one derived by ASCA. By
forcing the temperature at the value of \cite*{tsk00}, the
corresponding upper limits are $9.4\times 10^{-14}$ erg cm$^{-2}$
s$^{-1}$ ($L_X=1.1\times 10^{32}D^2_{3.2}$ \ergsec) and $8\times 10^8$
cm$^2$. While this large difference in the inferred black-body flux may
seem puzzling, it may be explained by the dramatic improvement of the
\xmm\  spatial resolution compared with the ASCA one. In fact, if we
compute the sum of our best-fit values for the flux of the black-body
component of the pl+bbody fit over all the rings and the ``edge"
region, we obtain a 0.5--10 keV flux of $1.2\times 10^{-12}$ erg
cm$^{-2}$ s$^{-1}$, or a luminosity of $1.3\times 10^{33}D^2_{3.2}$
\ergsec, in very good agreement with the value reported by H95 and
\cite*{tsk00}. More than 50\% of this flux comes from ring 7-8 and the
``edge" region: it is therefore probable that the ASCA data correctly
detected a soft excess in the integrated spectrum of 3C58, but this was
incorrectly attributed to the thermal component of the central compact
object, whereas Fig.  \ref{ratio} suggests that it comes from the {\em
external} parts of the nebula. The more stringent \xmm\  upper limit to
the black-body component also heavily constrains its interpretation.
Our value of the emission area is too low to assume that the whole
surface of neutron star is loosing the residual heat of formation, and
therefore would suggest the hot polar caps mechanisms. A review of the
polar caps heating mechanisms can be found by \cite*{yhh94}. H95 favors
the ``outer-gap" model, and in this case
we expect $L_X=10^{30}B_{12}P^{-2}$
\ergsec, where $B_{12}$ is the magnetic field in units of $10^{12}$ G
and P is the period in seconds. As shown in Fig.  \ref{h95rev}, our
$L_X$ upper limit yields a magnetic field higher than the highest value
known for allowed $B$ and $P$ values of an 810 yr old pulsar. For this
model to be compatible with the $L_X$ upper limit, either an older
pulsar must be present or, unlike the Crab, the pulsar has not suffered
high energy losses, in which case all the points below the isochrone
lines of Fig. \ref{h95rev} are allowed.  Other mechanisms invoked to
explain the hot polar caps yield estimates of $L_X$ lower then the
``outer-gap" model and in principle may be compatible with the
observations. However, other observed pulsar nebulae--thermal pulsar
pairs reported by \cite*{sw88} show a ratio between compact object
luminosity and total (pulsar+nebula) luminosity $L_{\rm c}/L_{\rm tot}$ of
$\sim 0.1-0.2$ (the Crab has $L_{\rm c}/L_{\rm tot}=0.04$), while, using the
observed total 3C58 luminosity of $1.8\times 10^{34}$ \ergsec and the
upper limit we have derived, we derive a ratio $<0.01$.  Fig.
\ref{h95rev} also shows that we expect a period less then 1 s for
reasonable values of $B$, and therefore, to properly detect a
pulsation, we would need a better time resolution than those of the
EPIC camera in full image mode.

\subsection{Thermal emission at the edge of the nebula}

\begin{figure}
  \centerline{\psfig{file=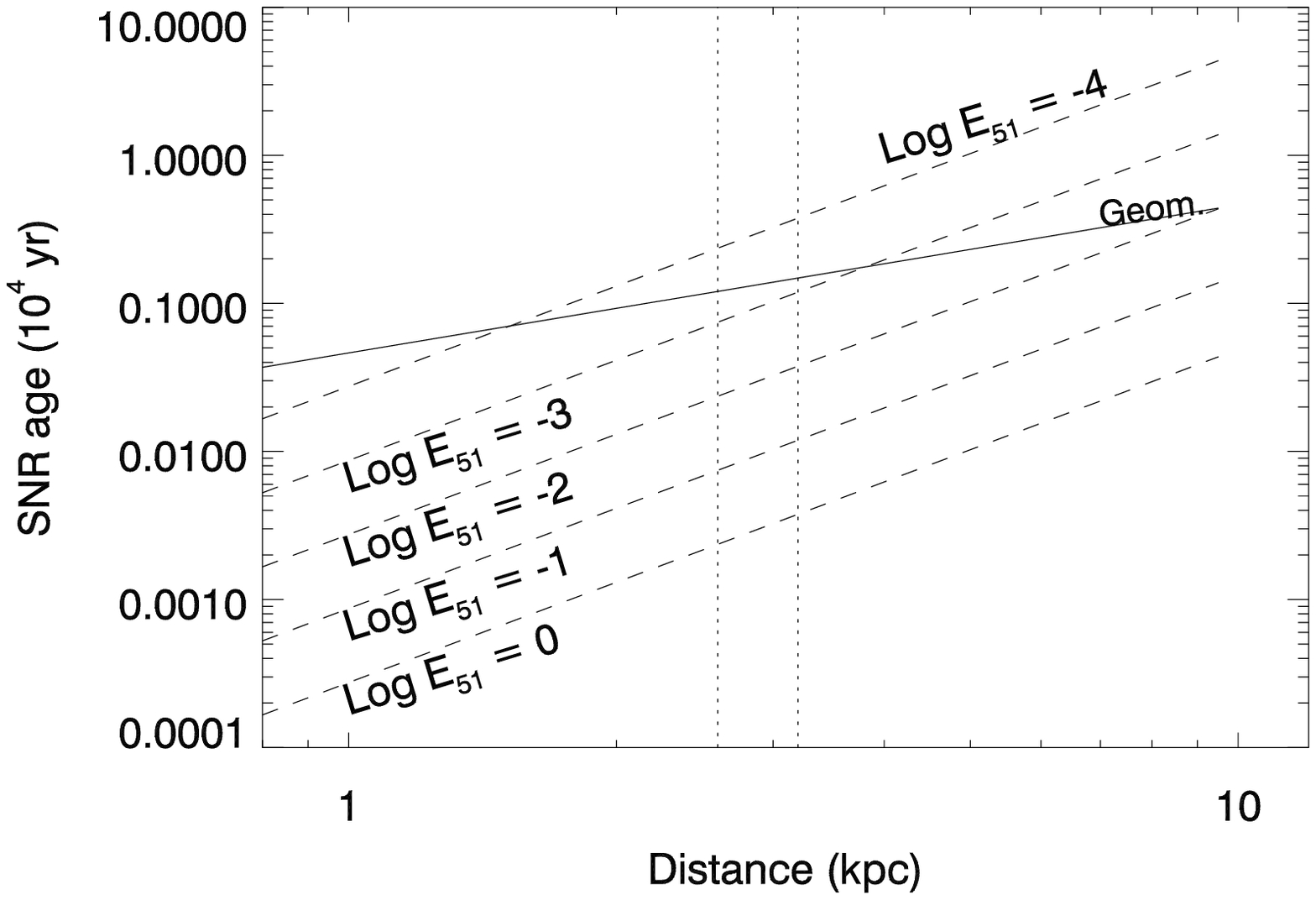,width=10cm}}
  \caption{Loci of allowed SNR age and distance according to a simple
  Sedov analysis of the thermal component of outer rim X-ray emission
  of 3C58. The solid line labeled ``Geom." gives the loci allowed by
  the simple geometrical relation between the real radius of the shell
  and the SNR age (Eq. 2 of \protect\cite{khw93}). The radius is linked
  to the distance via the apparent shell radius, and we have used
  2.5\arcmin\  to derive the relation. The dashed lines represent the
  loci of the solutions allowed by the Sedov relation
  $R=14(E_{51}/n_0)^{1/5} t_4^{2/5}$, where $n_0$ is derived from the
  emission measure of the {\sc mekal} component assuming a spherical
  emitting volume with a radius of 120\arcsec and a filling factor of
  25\%.  The intersection between the dashed lines and the solid line
  gives a solution.  Vertical dashed lines mark the distance estimate
  of 2.6 and 3.2, due to \protect\cite*{gg82} and
  \protect\cite*{rgk93}, respectively.}
  \label{sedov}
\end{figure}

The interpretation of the additional component in terms of expansion of
the main shock in the environment of the SNR cannot be excluded a
priori: recently, there have been some cases of SNRs reclassified as
composite (e.g. G11.2-0.3, \cite{vad96}; G327.1-1.1, \cite{swc99}).
The observed temperature of the {\sc mekal} component corresponds to a
shock speed of $\sim 450$ km s$^{-1}$, while the emissivity corresponds
to a post-shock density of $(0.60\pm 0.15)D_{3.2}^{-1/2}$ cm$^{-3}$ and
a pre-shock density four times smaller (assuming the emission comes
from a thin shell at $r=2.5\arcmin$, where the soft profile in Fig.
\ref{mos_prof} drops rapidly, and $\Delta r=r/12$). The emitting mass
is of the order of 0.1$D^{5/2}_{3.2}$ M$_\odot$. The inferred density
is very reasonable for typical low galactic latitude Sedov SNRs, but on
the other hand it can hardly be reconciled with the association between
SN1181 and 3C58. This is shown in Fig. \ref{sedov}, which reports the
allowed values of the SNR age and of the distance according to a simple
Sedov analysis following the outline given in \cite*{khw93}. It is
clear that from a pure geometrical point of view, the measured X-ray
temperature may represent a Sedov shock of an $\sim 800$ yr old SNR
located at a distance of 3.2 kpc (``Geom." solid line in Fig.
\ref{sedov}), but the measured emission measure of the {\sc mekal}
component yields a solution with a distance well above 10 kpc and and a
remnant age well in excess of $\sim 10^4$ yr, if we assume that the
explosion energy is of the order of the ``canonical" value $10^{51}$
erg ($\log E_{51}=0$ in Fig.  \ref{sedov}, where $E_{51}$ is in units
of $10^{51}$ erg).  Among the lowest values of the explosion energy
quoted in the literature, we have $E_{51}=0.1$ for Vela (\cite{bms99})
and $0.02-0.3$ for G292.0+1.8 (\cite{hs94}), and we note that, for
$E_{51}=0.01$, Fig.  \ref{sedov} gives a solution of a remnant at 8--10
kpc and an age of $3000-6000$ yr. The association 3C58-SN1181 suggests
an explosion energy of the order of $10^{48}$ erg, roughly ten times
lower than the lowest inferred.  We recall that the association have
been questioned by \cite*{hua86}, but recently \cite*{sg99} have
pointed out that, on the basis of an update of historical information,
the association should be reliable. If so, the Sedov model does not
provide a proper description of all the observational evidence.

In order to understand the interaction between 3C58 and its
environment, and to understand the origin of the soft X-ray component,
it is useful to compare its radio and X-ray emission.  The soft X-ray maps
of Fig. \ref{mos_imh} show that the X-ray emission is in any case
confined within 2\arcmin--4\arcmin\  from the center, and therefore the
presence of the shell is in agreement with the lack of detection of a
radio shell at $r>5\arcmin$ of \cite*{ra85}.  It is interesting to note
that there is a close correspondence between the radio morphology of
the outer regions of the nebula and the soft X-ray maps (Fig.
\ref{mos_imh}, upper panel). This is also observed in other shell-like
young SNR like Kepler (\cite{mld84}) and Cas A (\cite{kra96}), while,
on the other hand, the radio emission of the Crab nebula is four times
greater than its X-ray counterpart. \cite*{ra88} have pointed out that
the radio image of the remnant at 1446 MHz shows confined-edge
emission at some locations of the outer regions of the nebula, and some
of these also show limb-brightening. According to them, this may be
explained if the edge of the nebula is sweeping up moving
material, e.g. ejecta, like in the ``inhomogeneous" model of
\cite*{rc84}, or the shock model invoked by \cite*{sh97} to explain the
[O III] emission seen at the boundary of the Crab. This may explain the
small amount of limb brightening and is also in agreement with the
relatively slow shock speed measured in X-rays. In fact, the expected
shock speed in the moving ejecta material is $\sim 300$ km s$^{-1}$
(\cite{rc84}; 150-200 km s$^{-1}$ in the Crab according to
\cite{sh97}), and we observe $v_s=450$ km s$^{-1}$. Therefore, the
thermal component we observe in the outer rim of 3C58 may be associated
with the expansion of the nebula in the inner ejecta core. Note that
this is also in agreement with the discrepancy between the large speed
measured in 3C58 filaments ($\sim 900$ km s$^{-1}$, \cite{fes83}) and
the X-ray derived speed. In fact, the filaments are composed of
material ejected in the explosion itself, pushed on and accelerated by
the synchrotron nebula, while the X-ray emission is due a shock
expanding into moving ejecta.  However, the low value of the X-ray
emitting mass would imply that the interaction is only at its
beginning.

Finally, it should be noted that the above conclusions rely on the
hypothesis of fast electron-ion equipartition. Unfortunately, it is not
possible to independently measure the proton temperature $T_p$ of the 3C58
shock. \cite*{bms99} have shown that $T_e < T_p < 2T_e$ for a shock
region of the old Vela SNR, but \cite*{hrd00} found evidence for
$T_p\sim 45 T_e$ in the 1000 yr old SNR E0102.2-7219, suggesting that
non-equipartition may be common among young SNRs. If also in case of
3C58 $T_p$ is $\sim 45$ times higher than the X-ray derived electron
temperature, than a Sedov solution with $E_0 \sim 3\times 10^{49}$ erg,
would be compatible with a distance of 3.2 kpc and the association with
SN1181.

\section{Summary and conclusions}

The high spatial resolution of the EPIC camera on-board \xmm\  has
allowed us to look with unprecedented detail at the X-ray emission from
the filled-center SNR 3C58. We have presented soft (0.1--1 keV), medium
(1--2 keV) and hard (2--10 keV) energy images, and we have found a close
correlation of the soft image with radio emission at 1446 MHz. A
quantitative study of the source profile shows the effect of synchrotron
burn-off and indicates a different slope of the soft X-ray radial
profile compared to that obtained in the hard band. Spatially resolved
spectral analysis carried on in rings with $\Delta r=8\arcsec$ has
allowed us to derive the relation between the spectral index and the
distance from the core.  In contrast with previous ASCA results, we
have not found evidence for thermal black-body emission from a central
source, and we have placed an upper limit to this component of
$L_X=1.8\times 10^{32}$ \ergsec.  This stringent upper-limit rules out
thermal emission from the whole surface of the putative neutron star at
the center of 3C58, and also rules out the ``outer-gap" model for hot
polar caps of \cite*{chr86}, unless the pulsar has very low rotational
energy losses.

The large effective area of the \xmm\  mirrors has allowed us to
perform, for the first time, a spectral analysis of the outer edge of
the 3C58 X-ray nebula, excluding the inner and brighter regions. We
have found that non-thermal emission is still responsible for most of
the emission there, but a soft thermal component is required to better
fit the spectrum below 1 keV. If represented with an optically thin
emission model, this component gives kT=0.2--0.3 keV.  If it is
associated to the Sedov expansion of the 3C58 shell, it is incompatible
with the association between 3C58 and SN1181, unless there is strong
deviation from electron-ion equipartition.  If it is due to the
expansion of the nebula into the inner core of moving ejecta, the X-ray
spectra characteristics and the radio morphology of the outer nebula
can be more easily reconciled.

\begin{acknowledgements}

We thank the whole \xmm\  team for its effort in producing
well-calibrated data at this early stage of the mission.  F.~Bocchino
acknowledges an ESA Research Fellowship and thanks R. Bandiera for
helpful discussions.

\end{acknowledgements}

\bibliography{references}
\bibliographystyle{aabib}

\end{document}